\documentclass[11pt,aps,prd,preprint,groupedaddress,tightenlines,nofootinbib,floatfix]{revtex4}
\usepackage[dvipdfm]{graphicx}

\usepackage{amssymb}
\usepackage{amsmath}
\usepackage{epstopdf}
\usepackage{booktabs}

\usepackage{subfigure}
\makeatletter
\newcommand{\figcaption}[1]{\def\@captype{figure}\caption{#1}}
\newcommand{\tblcaption}[1]{\def\@captype{table}\caption{#1}}
\makeatother

\def\simge{\mathrel{%
       \rlap{\raise 0.511ex \hbox{$>$}}{\lower 0.511ex \hbox{$\sim$}}}}
\def\simle{\mathrel{
       \rlap{\raise 0.511ex \hbox{$<$}}{\lower 0.511ex \hbox{$\sim$}}}}

\begin{document}

\title{Phase structure of finite temperature QCD in the heavy quark region
\footnote{
We found an error in the analysis program of the effective potential developed for this paper after the publication, Phys.~Rev.~D {\bf 84}, 054502 (2011).
The error was in a coefficient of the term proportional to the plaquette. 
This causes slight shifts in the values of $\beta_{\rm trans}$ at $\kappa>0$ and thus $\beta_{\rm cp}$.
Accordingly, we have replaced FIG.~\ref{fig:dV}, FIG.~\ref{fig:beta_c}, TAB.~\ref{tab:Kcp_and_betacp} and Eq.~(\ref{eq:cp}) from those given in the published version [arXiv:1106.0974v2].
On the other hand, this error does not propagate to $d^2V_{\rm eff}/dP^2$. 
Therefore, the discussions and the conclusions of the paper, including the values of $\kappa_{\rm cp}$ as well as other figures and tables, are not affected. 
}}
\author{H.~Saito$^{1}$, S.~Ejiri$^{2}$, S.~Aoki$^{1,3}$, T.~Hatsuda$^{4,5,6}$, K.~Kanaya$^{1}$,  Y.~Maezawa$^{7}$, H.~Ohno$^{1}$\footnote{Present address: Physics Department, Bookhaven National Laboratory, Upton, New York 11973, USA.}
and T.~Umeda$^{8}$\\
(WHOT-QCD Collaboration)
}
\affiliation{$^1$Graduate School of Pure and Applied Sciences, University of Tsukuba, Tsukuba, Ibaraki 305-8571, Japan\\
$^2$Graduate School of Science and Technology, Niigata University, Niigata 950-2181, Japan\\
$^3$Center for Computational Sciences, University of Tsukuba, Tsukuba, Ibaraki 305-8577, Japan\\
$^4$Department of Physics, The University of Tokyo, Tokyo 113-0033, Japan\\
$^5$IPMU, The University of Tokyo, Kashiwa 277-8583, Japan\\
$^6$Theoretical Research Division, Nishina Center, RIKEN, Wako 351-0198, Japan\\
$^7$Mathmatical Physics Laboratory, Nishina Center, RIKEN, Wako 351-0198, Japan\\
$^8$Graduate School of Education, Hiroshima University, Hiroshima 739-8524, Japan
}
%\date{\today}
%\date{June 6, 2011}
%\date{August 11, 2011}
\date{February 29, 2012}

\begin{abstract}
We study the quark mass dependence of the finite temperature QCD phase transition in the heavy quark region using an effective potential defined through the probability distribution function of the average plaquette. 
Performing a simulation of SU(3) pure gauge theory, 
we first confirm that the distribution function has two peaks indicating that the phase transition is of first order in the heavy quark limit, while the first order transition turns into a crossover as the quark mass decreases from infinity, where
the mass dependence of the distribution function is evaluated by the reweighting method combined with the hopping parameter expansion.
We determine the endpoint of the first order transition region for $N_{\rm f}=1$, $2$, $3$ and $2+1$ cases. 
The quark mass dependence of the latent heat is also evaluated in the first order transition region. 
\end{abstract}

\maketitle

\section{Introduction}

The QCD phase transition is one of the important ingredients in understanding the evolution of the early universe. 
Monte-Carlo simulation of lattice QCD is the most powerful approach to study the nature of the QCD phase transition at present \cite{YHM}. 
It is known that the order of the phase transition depends on the values of quark masses:
The deconfinement phase transition is of first order when masses of all three (up, down and strange) quarks are either sufficiently large or small, 
while it turns into a crossover in the intermediate region \cite{Pisarski,Fukugita:1989yw,Iwasaki:1992ik,Ukawa,DeTar,Stickan01}.
Previous studies with staggered quarks strongly suggest that the transition becomes a crossover at physical quark masses \cite{Laermann03,Wupper2006,BNL-Bie2009}. 
A confirmation of this result by other fermion formulations such as Wilson-type fermions or by other methods of analyses is, however,  mandatory to draw a definite conclusion on the nature of the transition at physical quark masses \cite{iwasaki,cppacs1,whot09}.

Among several methods to study the nature of phase transitions, a probability distribution of physical observables provides us with the most intuitive way to determine the order of the phase transition. 
Adopting an appropriate physical quantity such as a quark number, chiral order parameter, gauge action etc., the corresponding probability distribution function is constructed by measuring 
a generation rate of configurations at each fixed value of the physical quantity.
Then an existence of double or multiple peaks in the probability distribution function give a signal of  a first order phase transition, since two phases coexist at a first order phase transition point.

In this study, we investigate the quark mass dependence of the order of the QCD phase transition in the large quark mass region using the hopping parameter expansion.
We expect the first order phase transition in the heavy quark limit becomes a crossover as we decrease the quark masses from infinity.
We take the plaquette, i.e. $1 \times 1$ Wilson loop, as the quantity to study the order of the transition. 
The reweighting technique \cite{Ferrenberg:1988yz,Ejiri:2007ga} 
is employed to vary quark masses in the lowest order of the hopping parameter expansion, and we determine an end point where the first order phase transition turns into a crossover. 

This paper is organized as follows. Basic properties of the plaquette
distribution function are discussed in Sec.~\ref{sec:plaquette}. 
A method to calculate the plaquette distribution function by the hopping parameter expansion is introduced in Sec.~\ref{sec:quarkdet}. 
Details of our simulations and results for the effective potential defined from the distribution function are presented in Sec.~\ref{sec:simulations}. 
Results show that the first order phase transition becomes a crossover as the quark mass decreases from infinity. 
We then determine the location of the end point  for the cases of $N_{\rm f}=1$, $2$, $3$ and $2+1$ with the Wilson quark action. 
Our conclusions are given in Sec.~\ref{sec:conclusion}.

\section{Probability distribution function}
\label{sec:plaquette}

A probability distribution function provides us with one of the most powerful methods to determine the order of phase transitions.
Since there exist two phases simultaneously at a first order phase transition point, 
we expect that the probability distribution function has multiple peaks near the transition point.
In this study, we consider the distribution function of the average plaquette $P$,
\begin{eqnarray}
P= \frac{1}{6 N_{\rm site}} \displaystyle \sum_{n,\mu < \nu} 
\frac{1}{3} {\rm Re \ tr} \left[ U_{n,\mu} U_{n+\hat{\mu},\nu}
U^{\dagger}_{n+\hat{\nu},\mu} U^{\dagger}_{n,\nu} \right],
\label{eq:plsaq}
\end{eqnarray}
where $U_{n,\mu}$ is the link variable and $N_{\rm site}=N_{\rm s}^3 \times N_t$ is the lattice volume. 
We adopt the standard one-plaquette action $(S_g)$ for glues and the standard Wilson fermion action $(S_q)$ for quarks:
\begin{eqnarray}
S_g &=& -6 N_{\rm site} \beta P, \\
S_q &=& \displaystyle \sum_{f=1}^{N_{\rm f}} \left\{ \sum_n\bar{\psi}_n^{(f)}\psi_n^{(f)} 
-\kappa_f \displaystyle \sum_{n,\mu}\bar{\psi}_n^{(f)} \left[ (1-\gamma_{\mu})U_{n,\mu}\psi_{n+\hat{\mu}}^{(f)}+(1+\gamma_{\mu})U_{n-\hat{\mu},\mu}^{\dagger}\psi_{n-\hat{\mu}}^{(f)} \right] \right\}\\
    & \equiv & \displaystyle \sum_{f=1}^{N_{\rm f}} \left\{ \sum_{n,m} \bar{\psi}_n^{(f)} M_{nm} (\kappa_f) \psi_m^{(f)}\right\} ,
\end{eqnarray} 
where $N_{\rm f}$ is the number of flavors, $\kappa_f$ is the hopping parameter for the $f$-th flavor, and $\beta = 6/g^2$ is the (inverse) gauge coupling.
The hopping parameter $\kappa_f$ controls 
the quark mass, which is proportional to $1/\kappa_f$ for small $\kappa_f$, 
while the lattice spacing is mainly controlled by $\beta$.
For the case of degenerate quark masses, i.e. $\kappa_f=\kappa$ for $f=1, \dots, N_{\rm f}$, 
the plaquette distribution function is defined as 
\begin{eqnarray}
w(P', \beta, \kappa) 
= \int {\cal D} U {\cal D} \psi {\cal D} \bar{\psi} \ \delta(P'-P) \ e^{- S_q - S_g}
= \int {\cal D} U \ \delta(P'-P) \ (\det M(\kappa))^{N_{\rm f}}
e^{6\beta N_{\rm site} P}. 
\label{eq:pdist}
\end{eqnarray}
The partition function is given by 
${\cal Z} (\kappa, \beta) = \int w(P', \beta, \kappa) dP' $.
In the followings, we denote the argument $P'$ in $w$ simply as $P$.

The plaquette distribution function has the following useful property \cite{Ejiri:2007ga}.
Under the change from $\beta_0$ to $\beta$, $w(P, \beta,\kappa)$ transforms as 
\begin{equation}
w(P,\beta,\kappa)=e^{6(\beta-\beta_0)N_{\rm site}P}w(P,\beta_0,\kappa).
\end{equation}
Therefore, the effective potential defined by
\begin{equation}
V_{\rm eff}(P,\beta,\kappa) = -\ln w(P,\beta,\kappa)
\end{equation}
transforms as
\begin{equation}
V_{\rm eff}(P, \beta, \kappa) 
= V_{\rm eff}(P, \beta_0, \kappa) - 6(\beta - \beta_0)N_{\rm site}P.
\label{eq:veffbeta}
\end{equation}
From this property, we find that 
\begin{eqnarray}
\frac{d V_{\rm eff}}{dP} (P,\beta, \kappa)
= \frac{d V_{\rm eff}}{dP} (P,\beta_0, \kappa) -6 (\beta - \beta_0) N_{\rm site},
\label{eq:derrewbeta}
\end{eqnarray}
and $d^2 V_{\rm eff}/dP^2$ is independent of $\beta$.

Since the distribution function is doubly peaked at and around a first order transition point, 
the corresponding effective potential has a double-well structure and the derivative  $d V_{\rm eff}/dP$ becomes an S-shaped function. 
Therefore, close to the transition point, $d V_{\rm eff}/dP$ must vanish at three points. 
In general, to find the first order phase transition by observing these properties, a fine tuning of $\beta$ is required. 
For the case of the plaquette effective potential, however,  we can detect the existence of the first order transition through the appearance of the S-shape in $d V_{\rm eff}/dP$ without fine tunings, since the $P$-dependence of $dV_{\rm eff}/dP$ itself is independent of $\beta$ as shown in Eq.~(\ref{eq:derrewbeta}). 
After the detection, the first order transition region in $\beta$, 
in which $d V_{\rm eff}/dP$ vanishes at three values of $P$, 
can be identified using Eq.~(\ref{eq:derrewbeta}).
Therefore, $d V_{\rm eff}/dP$ is useful to determine a region of the first order phase transition. 

We next discuss the $\kappa$-dependence of $V_{\rm eff}$ considering the ratio of the distribution functions at $\kappa$ and $\kappa_0$:
\begin{eqnarray}
R(P', \kappa, \kappa_0) 
& \equiv& \frac{w(P', \beta, \kappa)}{ w(P', \beta, \kappa_0)} \\
&=& \frac{\int {\cal D} U \delta(P'-P) (\det M(\kappa))^{N_{\rm f}} 
e^{6\beta N_{\rm site} P}}{\int {\cal D} U \delta(P'-P) 
(\det M(\kappa_0))^{N_{\rm f}} e^{6\beta N_{\rm site} P }} 
= \frac{\int {\cal D} U \delta(P'-P) (\det M(\kappa))^{N_{\rm f}}}{
\int {\cal D} U \delta(P'-P) (\det M(\kappa_0))^{N_{\rm f}}} \nonumber \\
&=& \frac{ \left\langle \delta(P'-P) 
\frac{(\det M(\kappa))^{N_{\rm f}}}{(\det M(\kappa_0))^{N_{\rm f}}} 
\right\rangle_{(\beta, \kappa_0)} }{
\left\langle \delta(P'-P) \right\rangle_{(\beta, \kappa_0)}}. 
\label{eq:rkdef}
\end{eqnarray}
Note that $R(P, \kappa, \kappa_0)$ is independent of $\beta$, and thus $R(P,\kappa, \kappa_0)$ can be evaluated at any $\beta$.
Using $R(P, \kappa, \kappa_0)$, the $\kappa$-dependence of the effective potential is given by 
\begin{equation}
V_{\rm eff}(P, \beta, \kappa) 
= -\ln R(P, \kappa, \kappa_0) + V_{\rm eff} (P,\beta, \kappa_0).
\label{eq:veff}
\end{equation}

This argument can be easily generalized to improved gauge actions including larger Wilson loops, 
for which we redefine the average plaquette as $P=-S_g/(6N_{\rm site}\beta)$. 
On the other hand, $\beta$-dependent improved quark actions make the analysis more complicated.

We note that the identification of the order of the phase transition by the $V_{\rm eff}$ is equivalent to that by the fourth order Binder cumulant, 
$%\begin{eqnarray}
B_4 = 
\left\langle (X- \langle X \rangle)^4 \right\rangle /
\left\langle (X- \langle X \rangle)^2 \right\rangle^2
$%\end{eqnarray}
with $X$ an appropriate operator signaling the phase transition,
since both $V_{\rm eff}$ and $B_4$ detect a change of the distribution function. 
On the other hand, to correctly evaluate $B_4$ at a first order transition point, we need to know precisely the probability distribution in a wide range of $X$ covering both peaks \cite{Ejiri:2007ga}.
A fine-tuning to the transition point and a high statistics with sufficiently many flip-flops are required for a reliable estimate of $B_4$ at the transition point. 
Furthermore, to achieve well separated peaks, a large system volume is required.
This makes the whole study quite demanding, in particular for weak first order transitions such as the case of the heavy quark limit of QCD.
On the other hand, a reliable $V_{\rm eff}$ in a wide range of $P$ can be easily obtained by combining data at different $\beta$ points thanks to the relation (\ref{eq:veffbeta}).

\section{Quark determinant in the heavy quark region}
\label{sec:quarkdet}

To investigate the quark mass dependence of the plaquette effective potential, we evaluate the quark determinant by a Taylor expansion with respect to the hopping parameter $\kappa$ in the vicinity of the simulation point $\kappa_0$:
\begin{eqnarray}
\ln \left[ \frac{\det M(\kappa)}{\det M(\kappa_0)} \right]
= \sum_{n=1}^{\infty} \frac{1}{n!} 
\left[ \frac{\partial^{n} (\ln \det M)}{\partial \kappa^{n}} 
\right]_{\kappa_0} (\kappa - \kappa_0)^{n} 
= \sum_{n=1}^{\infty} \frac{{\cal D}_{n}}{n!}  \, (\kappa - \kappa_0)^{n} , 
\label{eq:tayexp}
\end{eqnarray}
where 
\begin{eqnarray}
{\cal D}_n \equiv \left[ \frac{\partial^n \ln \det M}{\partial \kappa^n} \right]_{\kappa_0}
= (-1)^{n+1} (n-1)! \ {\rm tr} 
\left[ \left( M^{-1} \frac{\partial M}{\partial \kappa} \right)^n \right]_{\kappa_0}. 
\label{eq:derkappa}
\end{eqnarray}
Calculating the derivative of the quark determinant ${\cal D}_{n}$,
the $\kappa$-dependence of the effective potential can be estimated.

Adopting $\kappa_0=0$, the equation (\ref{eq:derkappa}) reads
\begin{eqnarray}
{\cal D}_n = (-1)^{n+1} (n-1)! \ {\rm tr} 
\left[ \left(\frac{\partial M}{\partial \kappa} \right)^n \right]_{\kappa=0}, 
\label{eq:derkappa0}
\end{eqnarray}
where $M_{x,y} = \delta_{x,y}$ at $\kappa=0$.
and $(\partial M/\partial \kappa)_{x,y}$ is the gauge connection between $x$ and $y$.
Therefore, the nonvanishing contributions to ${\cal D}_{n}$ are given by Wilson loops or Polyakov loops. Considering the anti-periodic boundary condition and gamma matrices in the hopping terms, the leading order contributions to the Taylor expansion are given by 
\begin{eqnarray}
\ln \left[ \frac{\det M (\kappa)}{\det M (0)} \right]  
=  288 N_{\rm site} \kappa^4 P + 12 \times 2^{N_t}N_s^3 \kappa^{N_t} {\rm Re}\Omega 
+ \cdots , 
\label{eq:detm}
\end{eqnarray}
where $\Omega$ is the Polyakov loop defined by
\begin{equation}
\Omega = \frac{1}{N_s^3}
\displaystyle \sum_{\mathbf{n}} \frac{1}{3} {\rm tr} \left[ 
U_{\mathbf{n},4} U_{\mathbf{n}+\hat{4},4} U_{\mathbf{n}+2\hat{4},4} 
\cdots U_{\mathbf{n}+(N_t -1)\hat{4},4} \right].
\label{eq:ploop}
\end{equation}
The ratio $R(P, \kappa, 0) = w(P, \beta, \kappa) / w(P, \beta, 0)$ 
is then calculated at arbitrary values of $\beta$ but small $\kappa$ as follows.
\begin{eqnarray}
R(P', \kappa, 0) 
&=& \frac{ \left\langle \delta(P'-P) \exp [N_{\rm f}
( 288 N_{\rm site} \kappa^4 P + 12 \times 2^{N_t} N_s^3\kappa^{N_t} {\rm Re}\Omega 
+ \cdots ) ] \right\rangle_{(\beta, \kappa_0=0)} }{
\left\langle \delta(P'-P) \right\rangle_{(\beta, \kappa_0=0)}} 
\nonumber \\
&=& e^{ 288 N_{\rm f}   N_{\rm site}\kappa^4 P' }
\frac{ \left\langle \delta(P'-P) \exp [N_{\rm f} 
(12 \times 2^{N_t} N_s^3 \kappa^{N_t} {\rm Re}\Omega 
+ \cdots ) ] \right\rangle_{(\beta, \kappa_0=0)} }{
\left\langle \delta(P'-P) \right\rangle_{(\beta, \kappa_0=0)}}. 
\label{eq:rewquench}
\end{eqnarray}
For the case of $N_t =4$ we study, the truncation error is $O(\kappa^6)$.
The contribution from the plaquette can be absorbed by a shift of $\beta$.

\section{Numerical simulations and the results}
\label{sec:simulations}

\begin{table}[t]
\centering
\begin{tabular}{ccc} \hline
$\beta$ & \ configurations \ & bin size \\
\hline
5.6800 & 100,000 &  100 \\
5.6850 & 430,000 & 2,150 \\
5.6900 & 500,000 & 2,000 \\
5.6925 & 670,000 & 3,350 \\
5.7000 & 100,000 &  500 \\
\hline
\end{tabular}
\caption{The number of configurations and the bin size for jackknife error analyses.}
\label{tab:beta_list}
\end{table}

\subsection{Simulation parameters and plaquette effective potential}

In the heavy quark limit we perform simulations of SU(3) pure gauge theory on a $24^3\times 4$ lattice.
We generate the configurations by a pseudo heat bath algorithm followed by four over-relaxation sweeps.
We perform simulations at five $\beta$ values in the range $5.68$ -- $5.70$. 
Simulation points and statistics are summarized in Table~\ref{tab:beta_list}.

\begin{figure}[t] 
   \begin{minipage}{7.5cm}
   \centering
   \includegraphics[width=7.2cm, clip]{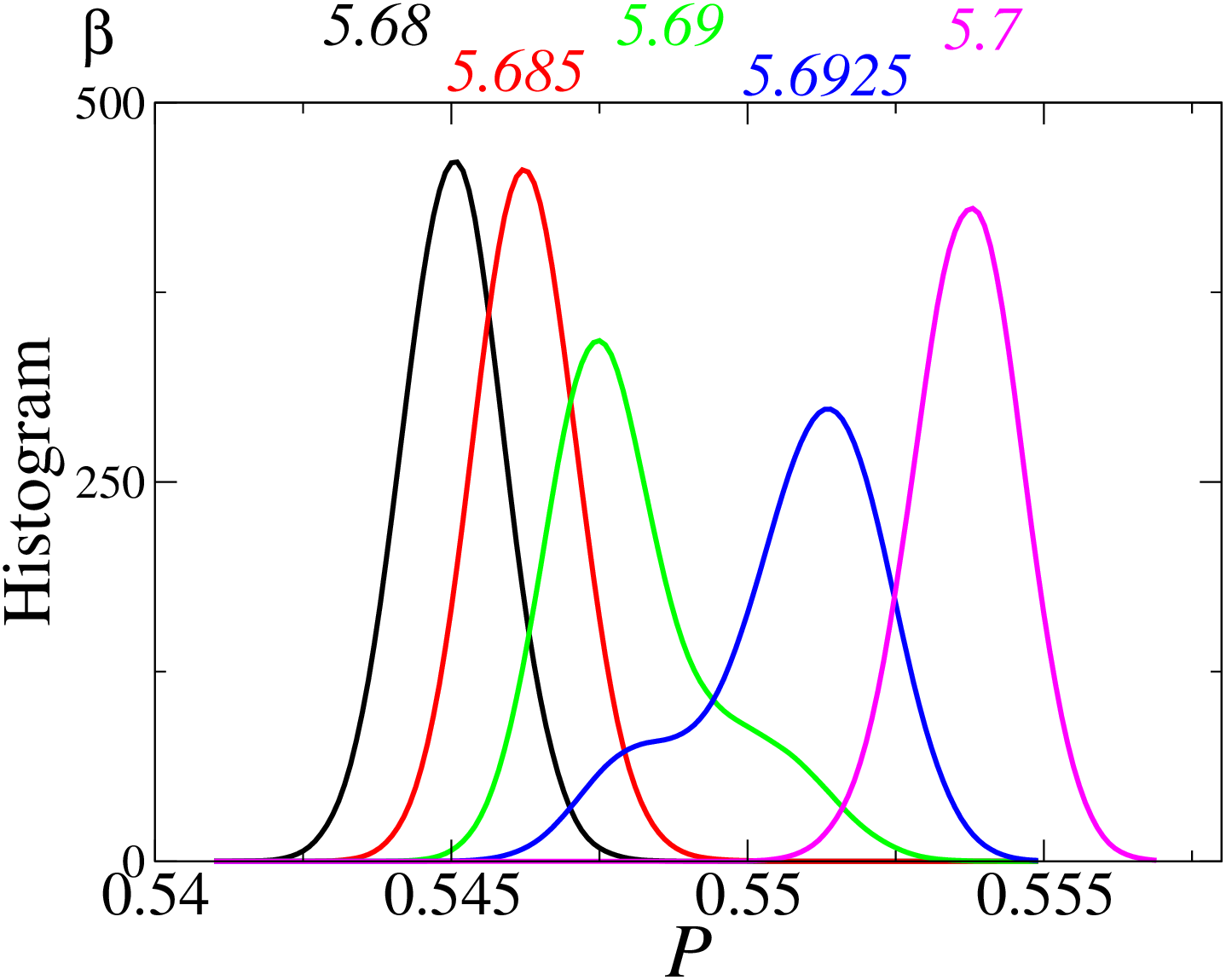} 
   \caption{Plaquette histogram $w/{\cal Z}$ obtained by SU(3) pure gauge simulations at $\beta=5.68$--$5.70$ on the $24^3 \times 4$ lattice. 
}
   \label{fig:hist}
   \end{minipage}
   \hspace{5mm}
   \begin{minipage}{7.5cm}
   \centering
   \includegraphics[width=7.5cm, trim=0 0 0 0, clip]{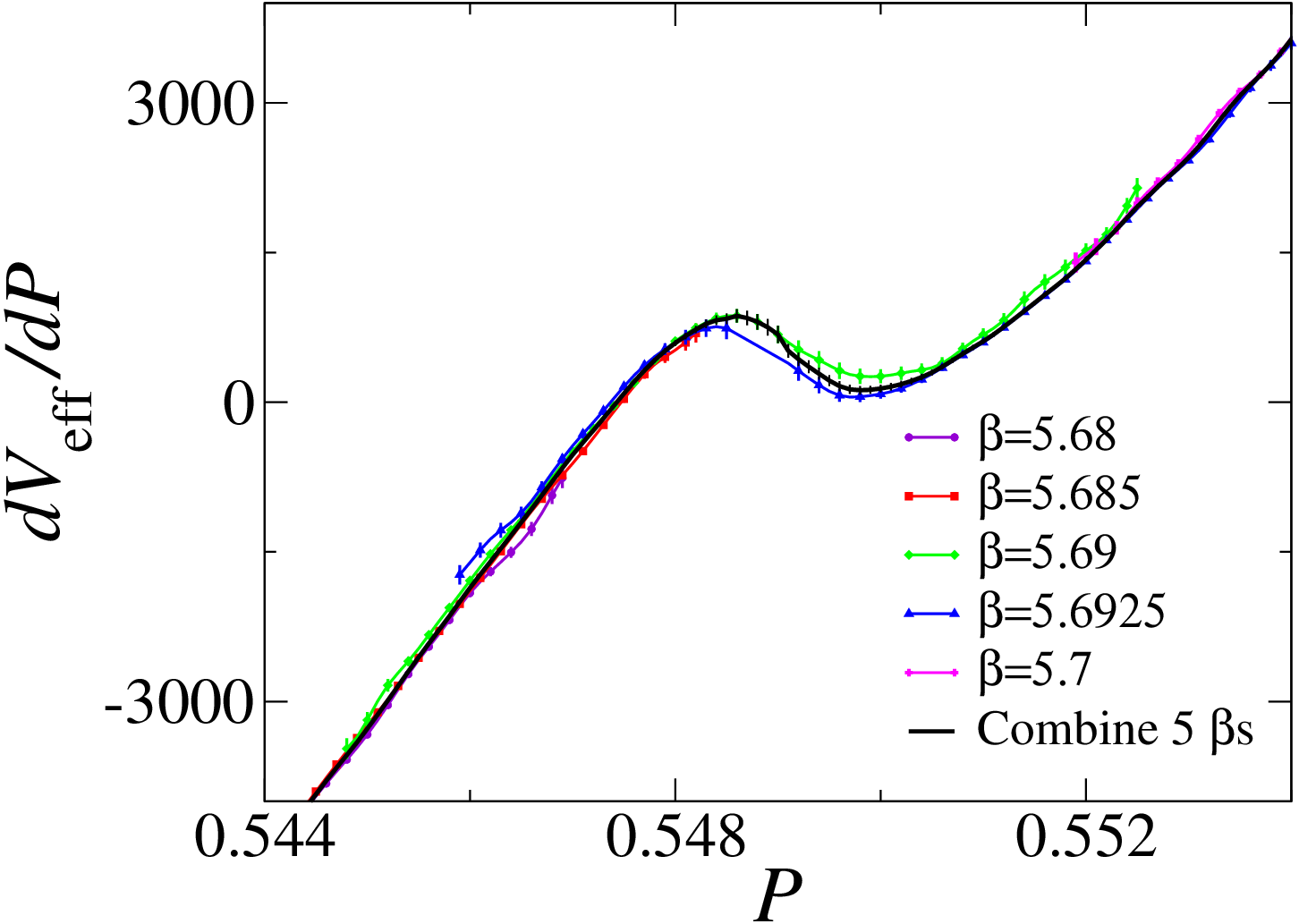} 
   \caption{Derivative of the effective potential in SU(3) pure gauge theory at $\beta=5.69$ converting data obtained at $\beta=5.68$--$5.70$ by Eq.~(\ref{eq:derrewbeta}). 
The bold black curve is an average over the results obtained at different $\beta$. }
   \label{fig:combine}
   \end{minipage}
\end{figure}

The results for the histogram of $P$, i.e. the plaquette distribution function Eq.~(\ref{eq:pdist}) normalized by the partition function ${\cal Z}$, are plotted in Fig.~\ref{fig:hist}. 
In this calculation, we approximate the delta function by a Gaussian function: 
$\delta(x) \approx 1/(\Delta \sqrt{\pi})\exp{ \left[-(x/\Delta)^2\right] }$, where $\Delta$ corresponds to the width of the Gaussian function. 
As $\Delta$ decreases, the resolution of the distribution function becomes better, while the statistical error increases. 
Examining the resolution and the statistical error, we adopt $\Delta = 0.000283$.
This figure shows that the value of $P$ generated in a simulation at single $\beta$ distributes in a narrow range. 

We then calculate the derivative $dV_{\rm eff}/dP$ by the difference between the potentials at $P - \epsilon_P/2$ and $P + \epsilon_P/2$. 
The value of $\epsilon_P$ is set to be $0.0001$, considering the resolution in $P$. The $\epsilon_P$-dependence is much smaller than the statistical error around this value of $\epsilon_P$.
Results of $dV_{\rm eff}/dP$ at $\kappa=0$ are shown in Fig.~\ref{fig:combine}, where data points with low statistics are removed for clarity. 
In this figure, we adjust results at different $\beta$'s to $\beta=5.69$ by using Eq.~(\ref{eq:derrewbeta}).
Results of $dV_{\rm eff}/dP$ obtained in simulations at different $\beta$ are consistent with each other within statistical errors, 
though the ranges of $P$ in which $V_{\rm eff}$ is reliably obtained are different.
The jackknife method is used to estimate the statistical error of the effective potential and its derivatives.
The bin size of the jackknife analysis is listed in Table \ref{tab:beta_list}. 
We then combine these data by taking an average weighted with the inverse-square of errors in the overlapping regions.
We here exclude data at $P$ far away from the peak of the distribution at each $\beta$ which thus has poor statistics. 
The final result for $dV_{\rm eff}/dP$ at $\kappa=0$ is shown by the black line without symbols in Fig.~\ref{fig:combine}.
We find that $dV_{\rm eff}/dP$ is not a monotonically increasing function at $\kappa=0$, 
indicating that the effective potential has a double-well shape.

\begin{figure}[t]
   \begin{minipage}{7.5cm}
      \includegraphics[width=7.5cm]{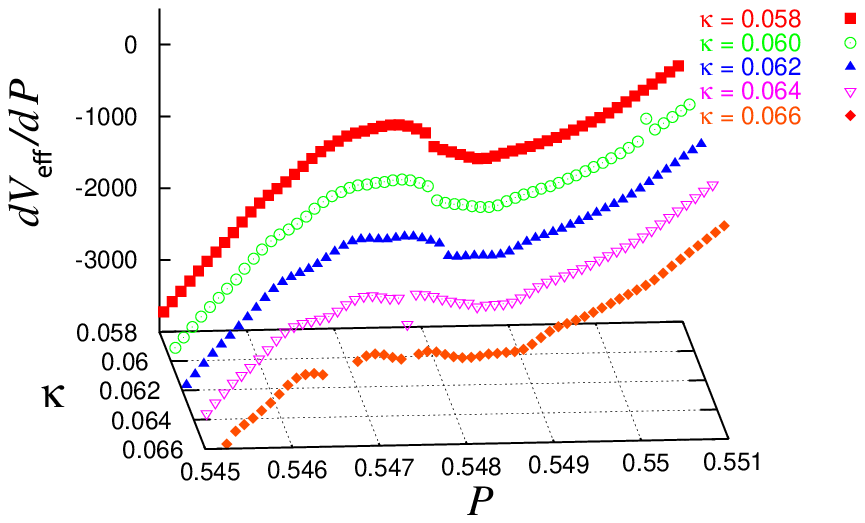}
   \caption{Derivative of the effective potential at nonzero $\kappa$ in two-flavor QCD.}
   \label{fig:dV}
   \end{minipage}
   \hspace{5mm}
   \begin{minipage}{8cm}
      \includegraphics[width=8cm, clip]{./K_V.eps}
   \caption{Effective potential $V_{\rm eff}$ at $\beta_{\rm trans}$ for each $\kappa$. }
   \label{fig:V}
   \end{minipage}
\end{figure}

The quark mass dependence of the effective potential is investigated by calculating $R(P, \kappa, 0)$ up to the order $\kappa^4$ in Eq.~(\ref{eq:rewquench}). 
Using the data of $w(P,\beta, \kappa=0)$ and $R(P, \kappa, 0)$, we evaluate the $dV_{\rm eff}/dP$ at nonzero $\kappa$. 
Results for $N_{\rm f}=2$ are plotted in Fig.~\ref{fig:dV}. 
The S-shape structure becomes weaker as $\kappa$ increases 
and turns into a monotonically increasing function around $\kappa = 0.066$. 
This behavior suggests that the first order phase transition at $\kappa=0$ becomes weaker as $\kappa$ increases 
and the transition becomes a crossover at $\kappa \approx 0.066$. 
In Fig.~\ref{fig:V} we show the $\kappa$ dependence of $V_{\rm eff}$ obtained by a numerical integration of $d V_{\rm eff}/ dP$. 
In this figure, $\beta$ is adjusted such that the two minima have the same height (see Sec.~\ref{sec:btrans} for details), and the integration is started from the peak point $P_{\rm peak}$ of $V_{\rm eff}$ between the two minima.
We find that the double-well becomes shallower as $\kappa$ increases and become almost flat around $\kappa = 0.066$.

\subsection{Critical point in the heavy quark region for $N_{\rm f}=2$}
\label{sec:criticalpoint}

In this subsection, we evaluate the value of $\kappa$ at which the first order phase transition terminates. 
We denote the corresponding critical point as $\kappa_{\rm cp}$.

\subsubsection{Critical point from $d^2 V_{\rm eff}/dP^2$}

We first calculate the second derivative of $V_{\rm eff}(P, \beta, \kappa)$ by a numerical differentiation of $dV_{\rm eff}/dP$.
When the transition is of first order, there is a region in $P$ where $d^2 V_{\rm eff}/dP^2$ is negative between the two bottoms of $V_{\rm eff}(P, \beta, \kappa)$. 
The first order transition region is thus identified by calculating the sign of the second derivative of $V_{\rm eff}$. 
As discussed in Sec.~\ref{sec:plaquette}, the second derivative of $V_{\rm eff}(P, \beta, \kappa)$ is independent of $\beta$. 
Therefore, the identification can be performed at any $\beta$.

\begin{figure}[t]
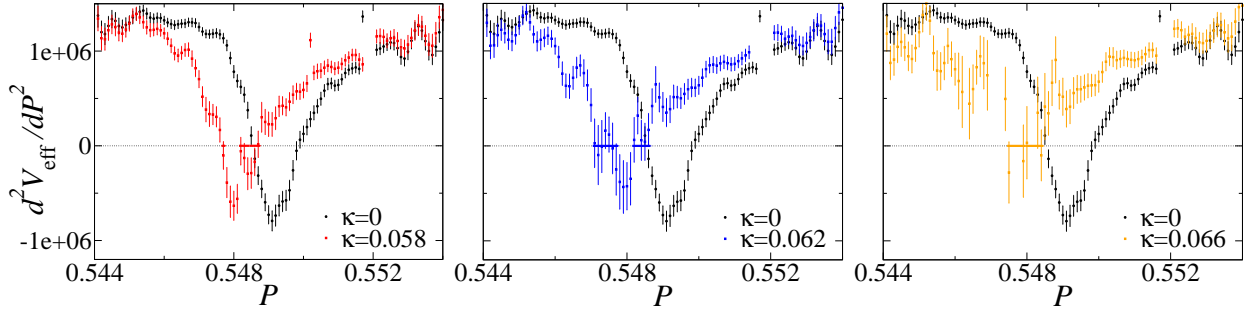
 
  \begin{minipage}{5.8cm}
     \includegraphics[width=5.8cm,  clip]{./ddV_K0058.eps}
   \end{minipage} 
   \begin{minipage}{5.2cm}
     \includegraphics[width=5.2cm, clip]{./ddV_K0062.eps}
   \end{minipage}
   \begin{minipage}{5.2cm}
     \includegraphics[width=5.2cm, clip]{.//ddV_K0066.eps}
   \end{minipage}
   \caption{Curvature of the effective potential at $\kappa=0.058$ (left), $0.062$ (middle), $0.066$ (right). 
Black symbols are results at $\kappa=0$.
Bold horizontal lines on the line of $d^2V_{\rm eff}/dP^2=0$ represent errors of the plaquette values where $d^2V_{\rm eff}/dP^2$ vanish.
}
   \label{fig:ddV}
\end{figure}

\begin{figure}[t]
   \begin{minipage}{7cm}
      \includegraphics[width=6.8cm, clip]{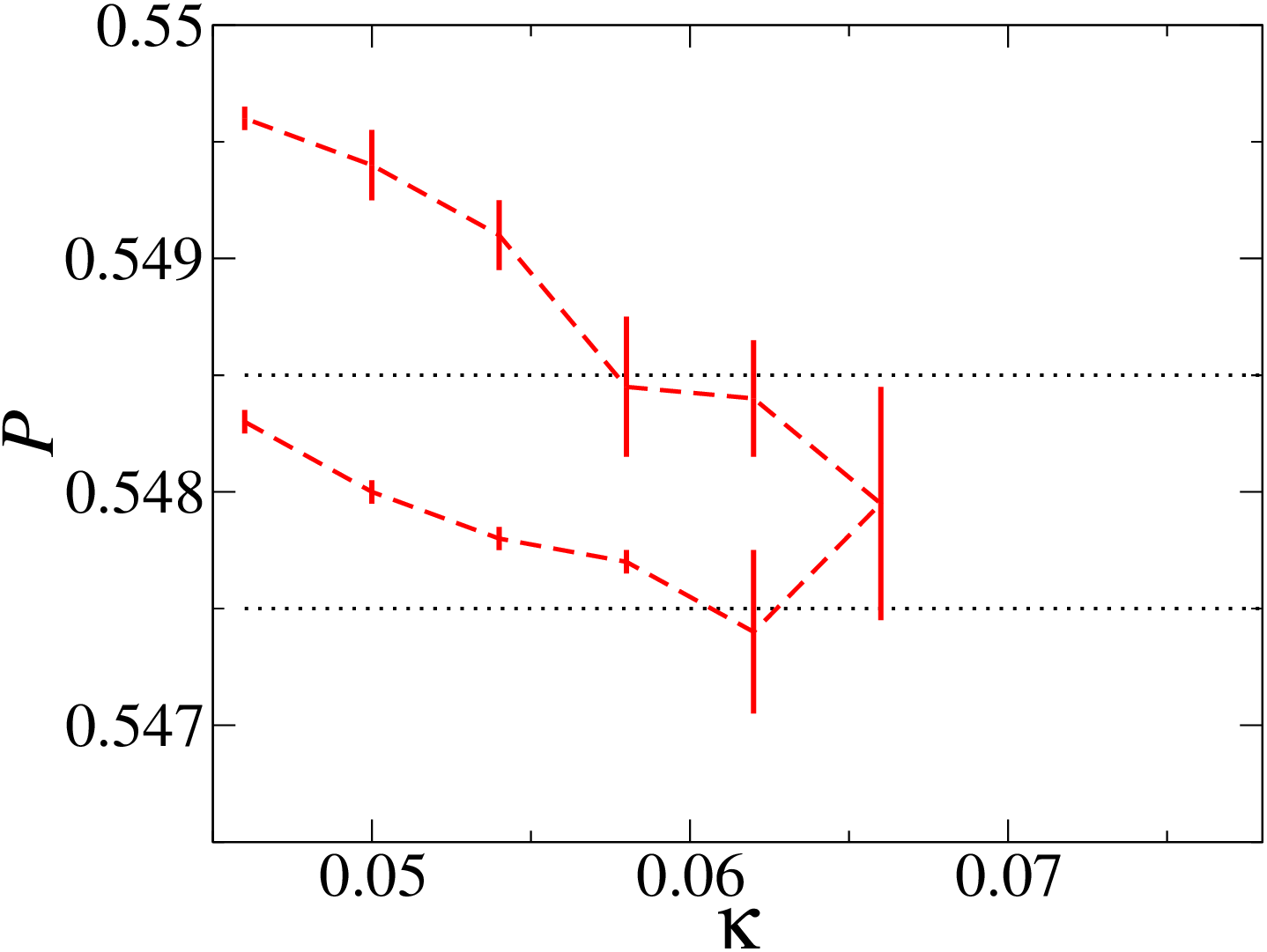}
   \end{minipage}
   \caption{Value of $P$ where $d^2V_{\rm eff}/dP^2$ vanishes at each $\kappa$. }
   \label{fig:ddV0_plaq_earlyestimate}
\end{figure}

In Fig.~\ref{fig:ddV}, we plot the results of $d^2 V_{\rm eff}/dP^2$ at $\kappa=0.058$ (left), $0.062$ (middle) and $0.066$ (right) for $N_{\rm f}=2$, together with the results at $\kappa=0$ which are shown as black symbols. 
We find that the region where $d^2 V_{\rm eff}/dP^2 < 0$ becomes narrower as $\kappa$ increases. 
Bold horizontal lines at zero %on the line of $d^2 V_{\rm eff}/dP^2 = 0$ 
show the range of $P$ where the curvature vanishes within statistical errors.
Hereafter we denote the position at which $d^2 V_{\rm eff}/dP^2 = 0$ as 
$P_{(V_{\rm eff}''=0)} (\kappa)$ or $\kappa_{(V_{\rm eff}''=0)} (P)$. 
Results of $P_{(V_{\rm eff}''=0)}$ are plotted in 
Fig.~\ref{fig:ddV0_plaq_earlyestimate} as a function of $\kappa$. 
The region of the negative curvature seems to vanish around $\kappa = 0.066$. 
Since the curvature is always positive beyond this $\kappa$, the effective potential is no longer a double-well type, i.e. the transition becomes a crossover at  $\kappa \simge 0.066$.

\begin{figure}[t]
   \begin{minipage}{7.9cm}
   \begin{minipage}{3cm}
      \centering
      \includegraphics[width=3cm, clip]{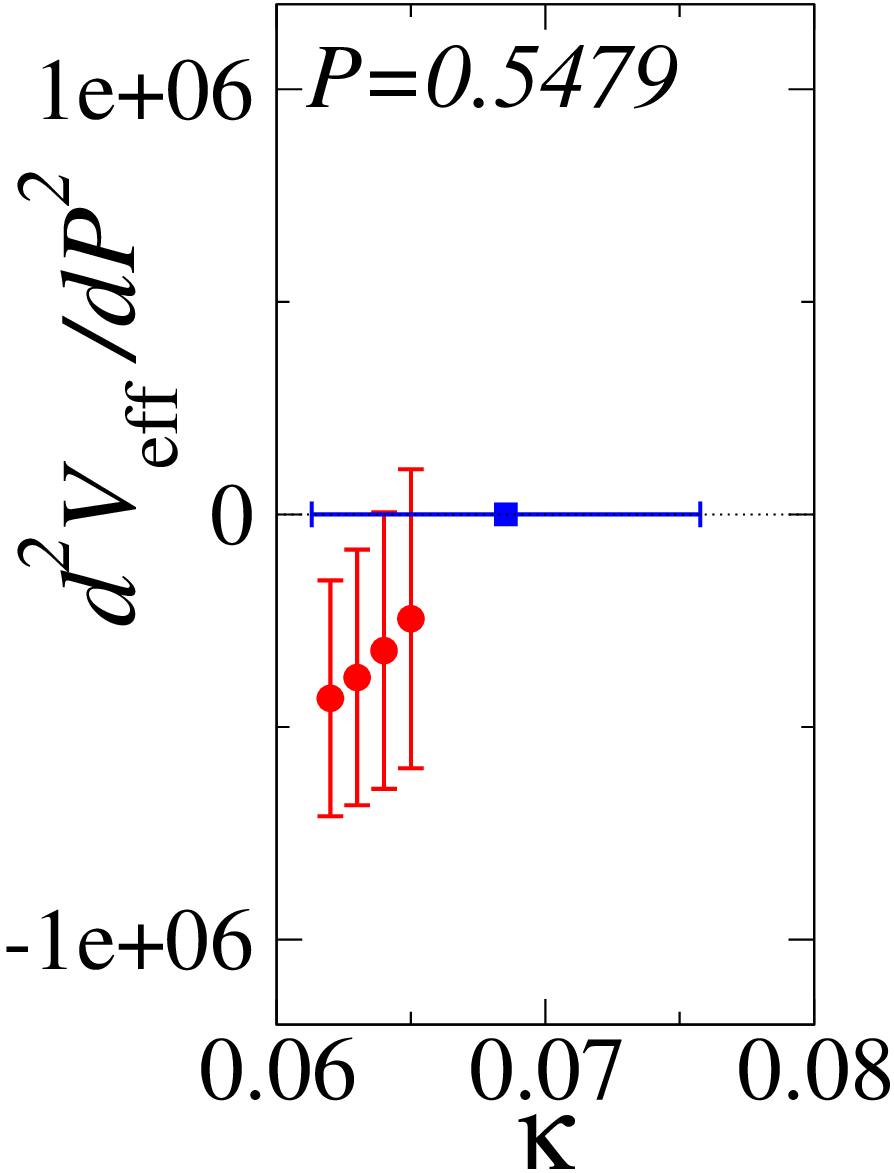}
   \end{minipage}
   \hspace{0.5mm}
   \begin{minipage}{1.85cm}
      \centering
      \includegraphics[width=1.85cm, clip]{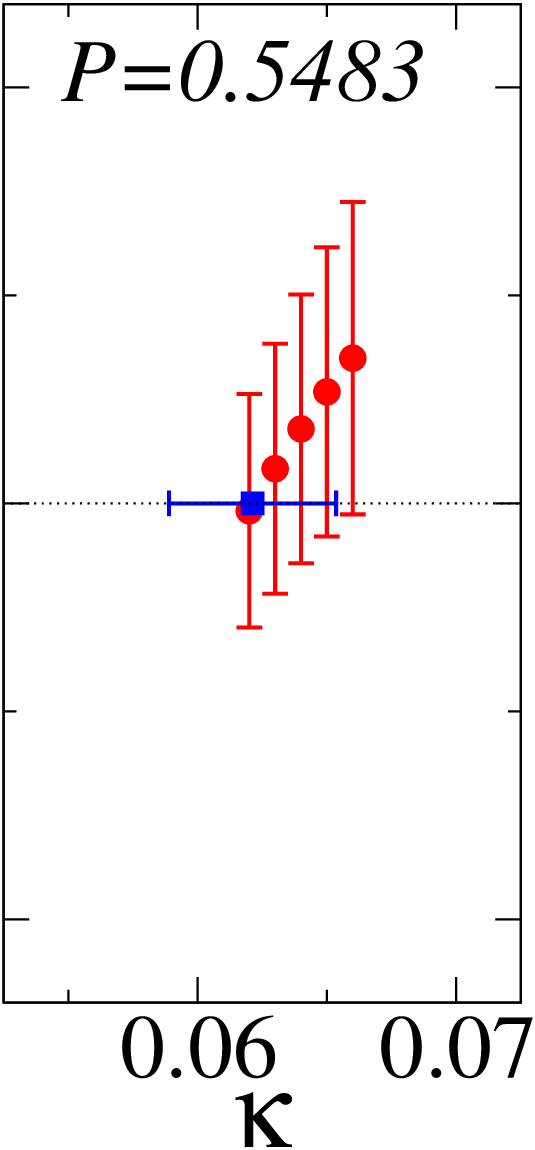}
   \end{minipage}
   \hspace{0.5mm}
   \begin{minipage}{2.3cm}
      \centering
      \includegraphics[width=2.3cm, clip]{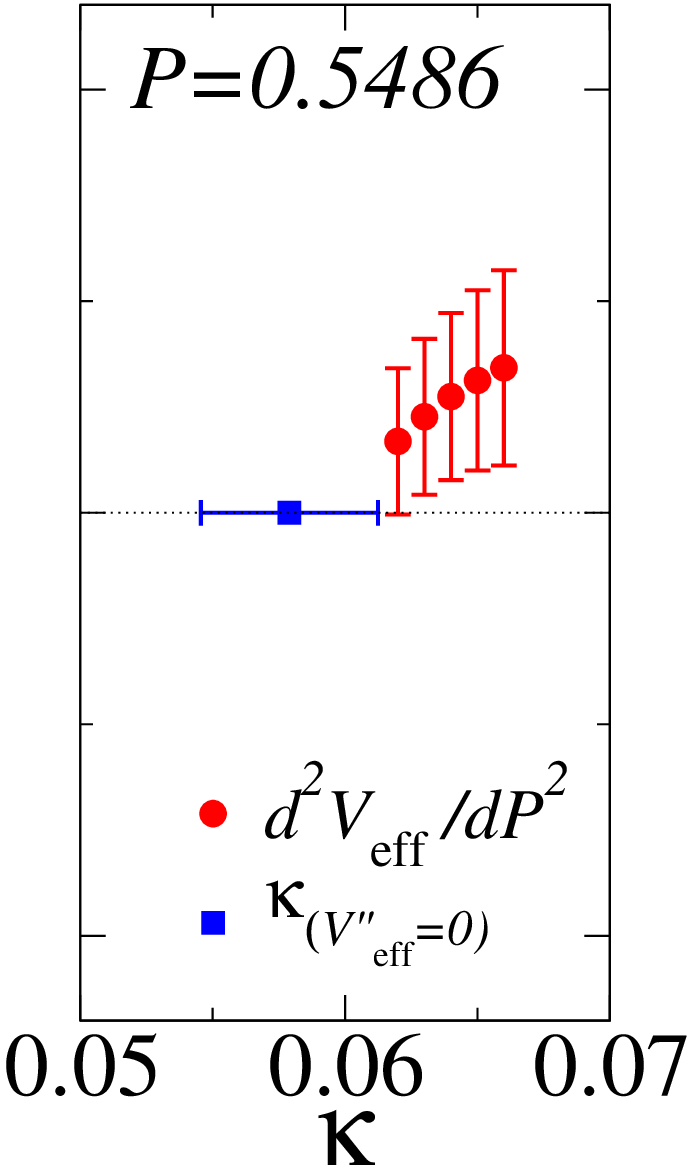}
   \end{minipage}
   \caption{$\kappa$-dependence of $d^2V_{\rm eff}/dP^2$ and the results of $\kappa_{(V_{\rm eff}''=0)}$ at $P=0.5479$, $0.5483$ and $0.5486$.}
   \label{fig:ddV0_plaq_fit}
   \end{minipage}
   \hspace{3mm}
   \begin{minipage}{7.5cm}
      \centering
      \includegraphics[width=7cm, clip]{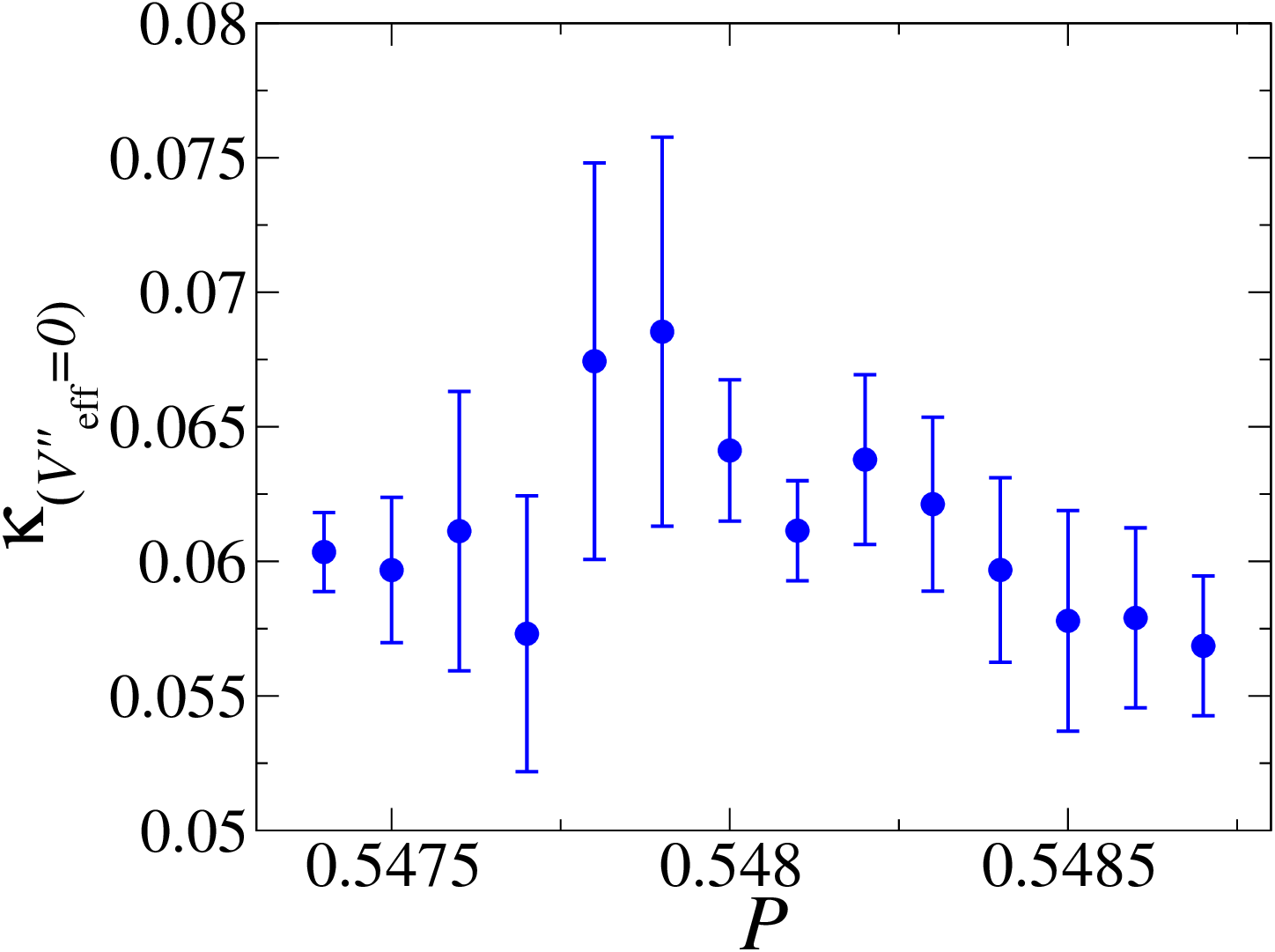} 
      \caption{Value of $\kappa$ where $d^2V_{\rm eff}/dP^2$ vanishes at each $P$.}
      \label{fig:ddV0_plaq}
   \end{minipage}
\end{figure}

To evaluate the critical point $\kappa_{\rm cp}$ we study the $\kappa$-dependence of $d^2 V_{\rm eff}/dP^2$ at fixed $P$, as shown in Fig.~\ref{fig:ddV0_plaq_fit} for typical points.
Data with large statistical errors are removed from this figure. 
Fitting data by a linear function in $\kappa$, we obtain $\kappa_{(V_{\rm eff}''=0)(P)}$, which is shown by a square symbol in the same figure.
The horizontal bar represents the statistical error.
Results of $\kappa_{(V_{\rm eff}''=0)}$ are plotted in Fig.~\ref{fig:ddV0_plaq} for $0.5474 \le P \le 0.5487$. 
When $\kappa$ is larger than the maximum value of $\kappa_{(V_{\rm eff}''=0)}$, $d^2V_{\rm eff}/dP^2$ is non-negative for all values of $P$ in this region.
Therefore, the maximum value of $\kappa_{(V_{\rm eff}''=0)}$ is nothing but the critical point $\kappa_{\rm cp}$.
We obtain $\kappa_{\rm cp}=0.0685(72)$ for $N_{\rm f}=2$.

\subsubsection{Critical point from the double-well structure of $V_{\rm eff}$}
\label{sec:btrans}

\begin{figure}[t]
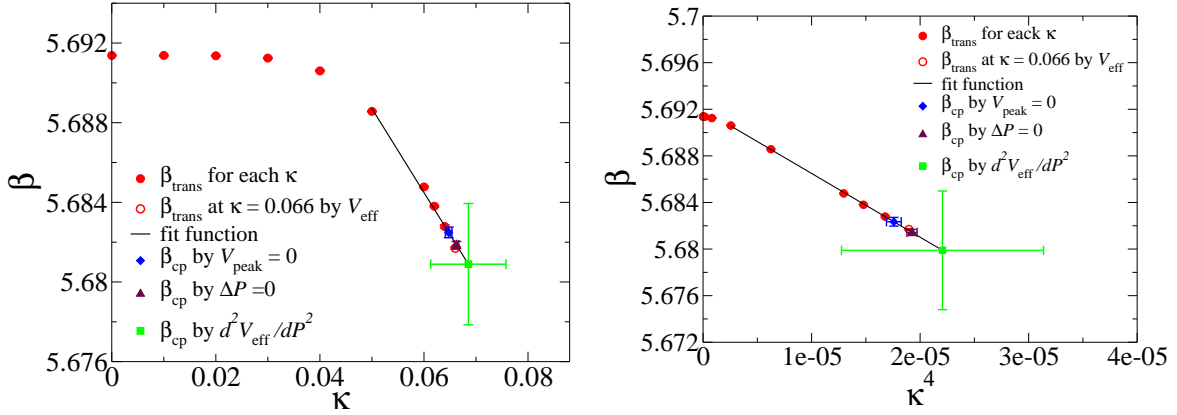

   \begin{minipage}{7.5cm}
   \centering
   \includegraphics[width=7.5cm, clip]{./beta_c_g.eps} 
   \end{minipage}
   \hspace{2mm}
   \begin{minipage}{7.5cm}
   \centering
   \includegraphics[width=7.5cm, clip]{./beta_c_K4_g.eps} 
   \end{minipage}
   \caption{$\beta_{\rm trans}$ as a function of $\kappa$ (Left) and $\kappa^4$ (Right) for $N_{\rm f}=2$. 
    Also shown are the results of the critical point $(\beta_{\rm cp})$, which are obtained by linearly extrapolating $\beta_{\rm trans}$ in $\kappa$ (Left) or $\kappa^4$ (Right) to $\kappa_{\rm cp}$ determined by $V_{\rm peak}$ (diamonds), $\Delta P$(triangles), or $d^2V_{\rm eff}/dP^2$ (squares).}
\label{fig:beta_c}
\end{figure}

Alternatively, we may determine $\kappa_{\rm cp}$ as the point where the two minima of $V_{\rm eff}(P)$ merge and the barrier between the two minima vanishes.
The double-well structure of $V_{\rm eff}$ is most clearly seen at $\beta$ where the two minima of $V_{\rm eff}$ has the same height (see Fig.~\ref{fig:V}).
Such $\beta$, say $\beta_{\rm trans}$, can be determined by a Maxwell's construction for $dV_{\rm eff}/dP$:
Let us denote the values of $P$ at the two minima as $P_{\rm A}$ and $P_{\rm B}$. 
The condition $V_{\rm eff}(P_{\rm A})=V_{\rm eff} (P_{\rm B})$ implies 
$ %\begin{equation}
{\displaystyle
\int_{P_{\rm A}}^{P_{\rm B}} \frac{d V_{\rm eff}}{dP} (P,\beta_{\rm trans}, \kappa) \, dP =0
}
$, %\end{equation}
or, equivalently,
\begin{equation}
\int_{P_{\rm A}}^{P_{\rm peak}} \frac{d V_{\rm eff}}{dP} (P,\beta_{\rm trans}, \kappa) dP
= - \int_{P_{\rm peak}}^{P_{\rm B}} \frac{d V_{\rm eff}}{dP} (P,\beta_{\rm trans}, \kappa) dP
\label{eq:intbc}
\end{equation}
where $P_{\rm peak}$ is the peak position of $V_{\rm eff}$ between $P_{\rm A}$ and $P_{\rm B}$ at which 
$dV_{\rm eff}/dP$ vanishes.
Changing $\beta$ by Eq.~(\ref{eq:derrewbeta}), we search $\beta_{\rm trans}$ where Eq.~(\ref{eq:intbc}) is satisfied.
The results for $\beta_{\rm trans}$ are plotted in the left panel of Fig.~\ref{fig:beta_c} as a function of $\kappa$. 
At $\kappa=0$, we obtain $\beta_{\rm trans}=5.69138(3)$. 
At $\kappa=0.066$, because the minima of $V_{\rm eff}$ are too shallow to apply the Maxwell construction, we instead determine $\beta_{\rm trans}$ as the point where the bottom region of $V_{\rm eff}$ becomes flat, as shown in Fig.~\ref{fig:V}.
In the right pane of Fig.~\ref{fig:beta_c}, we replot $\beta_{\rm trans}$ as a function of $\kappa^4$, as motivated by the hopping parameter expansion.
We note that the data is well fitted by a linear function of $\kappa^4$.

\begin{figure}[t]
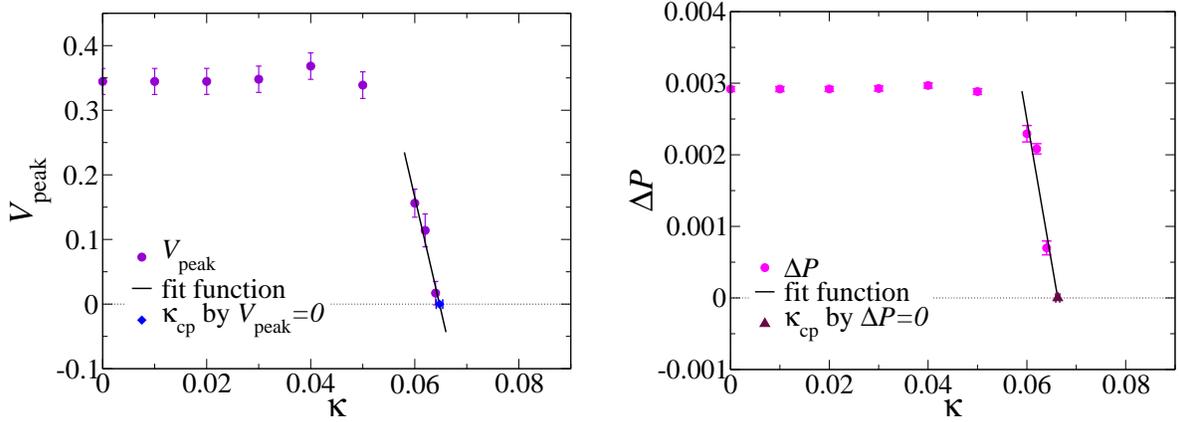

   \begin{minipage}{16cm}
   \centering
   \includegraphics[width=7.5cm, clip]{./height_peak.eps} 
   \hspace{5mm}
   \includegraphics[width=7.3cm, clip]{./delta_p.eps} 
   \caption{Potential barrier height $V_{\rm peak}$ (Left) and the gap $\Delta P$ (Right) between the two minima at $\beta_{\rm trans}$ for $N_{\rm f}=2$. Critical point $\kappa_{\rm cp}$ is estimated by linear extrapolations shown in the figures.}
   \label{fig:height_peak}
   \end{minipage}
\end{figure}

In an approximation to disregard fluctuations around the plaquette expectation value $P$, we can identify $V_{\rm eff}(P)$ with the free energy of the system. 
Then the condition $V_{\rm eff}(P_{\rm A})=V_{\rm eff} (P_{\rm B})$ means that the pressure $p$ is balanced between the phases A and B.
Therefore, we may view $\beta_{\rm trans}$ as an estimate of the first order transition point.
Actually, our result $\beta_{\rm trans}=5.69138(3)$ for $\kappa=0$ is quite close to the first order transition point 
$\beta_{\rm trans}=5.69153(86)$--5.69211(35) defined by the peak position of the plaquette susceptibility in SU(3) pure gauge theory with the same lattice size $24^3\times4$ \cite{Fukugita:1989yw}.

In Fig.~\ref{fig:height_peak}, results of the potential barrier height $V_{\rm peak} \equiv V_{\rm eff}(P_{\rm peak}) - V_{\rm eff}(P_{\rm A})$ and the gap $\Delta P \equiv P_{\rm B} - P_{\rm A}$ at $\beta_{\rm trans}$ are shown as functions of $\kappa$.
We find that both $V_{\rm peak}$ and $\Delta P$ decrease drastically above $\kappa = 0.05$ and vanish around $\kappa \approx 0.066$.
Performing a linear extrapolation using three nearest points of $V_{\rm peak}$ [$\Delta P$] as shown in Fig.~\ref{fig:height_peak}, we obtain $\kappa_{\rm cp} = 0.0647(6)$ [$0.0662(4)$].
A corresponding value of $\beta_{\rm cp}$ at the critical point can be estimated by extrapolating $\beta_{\rm trans}$ to $\kappa_{\rm cp}$,
assuming a linear function in $\kappa$ or $\kappa^4$ as  plotted in Fig.~\ref{fig:beta_c}. 
Results for $\kappa_{\rm cp}$ and $\beta_{\rm cp}$are summarized in Table~\ref{tab:Kcp_and_betacp}.
In the table, we also present $\beta_{\rm cp}$ for  $\kappa_{\rm cp}$ determined by $d^2 V_{\rm eff}/dP^2$.
From these results, we obtain
\begin{eqnarray}
\kappa_{\rm cp} = 0.0658(3)(^{+4}_{-11}),
\hspace{1cm}
\beta_{\rm cp} = 5.6819(1)(5)
\label{eq:cp}
\end{eqnarray}
where the central values and their statistical errors given in the first brackets are determined by a weighted average of the results, 
and systematic errors given in the second brackets are 
determined from the scattering of the results due to the method and extrapolation function, neglecting the data from the $d^2V_{\rm eff}/dP^2$ method which are consistent with other results within large errors.

To get a rough idea about the value of the pseudoscalar meson mass $(m_{\rm PS})$ corresponding to the critical point, 
we estimate the ratio $T/m_{\rm PS} = 1/(N_t m_{\rm PS} a)$ at the critical point.
Using a data of zero-temperature pseudoscalar meson mass in $N_{\rm f}=2$ QCD along the finite temperature crossover curve for $N_t=4$ in the range $\kappa=0.16$--0.19 \cite{Bitar91}, 
we perform a fit $(m_{\rm PS}a)^{-1} = f_1 \kappa^2 + f_2 \kappa^4$ with fitting parameters $f_1$ and $f_2$, 
where the constraint $(m_{\rm PS}a)^{-1} =0$ at $\kappa=0$ is taken into account.
We find $T/m_{\rm PS} \approx 0.023$ for $N_{\rm f}=2$ at the critical point $\kappa_{\rm cp} \approx 0.066$. 
Because $T_{\rm trans}/m_{\rm PS} = O(1)$ around the physical point, the small value 0.023 means that the critical pseudoscalar meson mass is much larger than the physical pion mass. 

\begin{table}[h]
   \setcounter{table}{1}
   \centering
   \begin{tabular}{cccc} \hline 
                                                  &  $\kappa_{\rm cp}$     & \multicolumn{2}{c}{$\beta_{\rm cp}$} \\ \cline{3-4}
       method                             &                                      & $\kappa$ fit         & $\kappa^4$ fit\\ \hline
      $V_{\rm peak}$                & $0.0647(06)$               & $5.6824(02)$     & $5.6823(03)$ \\ 
      $\Delta P$                         & $0.0662(04)$               & $5.6818(01)$     & $5.6814(02)$ \\ 
      $d^2V_{\rm eff}/dP^2$    & $0.0685(72)$            & $5.6808(30)$     & $5.6798(50)$ \\
      \hline
      total                   &    $0.0658(03)\left( ^{+4}_{-11}\right) $             & \multicolumn{2}{c}{$5.6819(1)(5)$} \\
   \hline
   \end{tabular}
\caption{Critical point $\kappa_{\rm cp}$ and $\beta_{\rm cp}$ defined by $V_{\rm peak}$, $\Delta P$ and $d^2V_{\rm eff}/dP^2$}
\label{tab:Kcp_and_betacp}
\end{table}

\subsection{Latent heat}

The gap of the internal energy density, $\Delta \varepsilon$, at a first order transition point is called the latent heat. 
Since $p$ is continuous there, $\Delta \varepsilon$ can be evaluated as the gap of the trace anomaly: 
\begin{eqnarray}
\frac{\varepsilon - 3p}{T^4}
&=& -\frac{1}{VT^3} a \frac{d \ln {\cal Z}}{d a} 
\,-\,[T=0]
% = \frac{N_t^3}{N_s^3} \frac{a}{\cal Z} \frac{d {\cal Z}}{d a} -[T=0] 
\nonumber \\
%&=& - \frac{N_t^3}{N_s^3} \frac{1}{\cal Z} \int {\cal D} U \left[ N_{\rm f} \left( a \frac{d \kappa}{d a} 4 \kappa^3 \cdot 288 N_s^3 N_t P 
%+  a \frac{d \kappa}{d a} \kappa^{N_t-1} \cdot 12 \times 2^{N_t} N_s^3 {\rm Re}\Omega \right) \right. \nonumber \\ && \left.
%+ a \frac{d \beta}{d a} 6 N_s^3 N_t P + O(\kappa^6) \right] (\det M)^{N_{\rm f}} e^{-S_g} -[T=0] \nonumber \\
&=& - N_t^4 \left[
a \frac{d \beta}{d a} 6 \langle P \rangle
+ N_{\rm f} a \frac{d \kappa}{d a} \left( 1152 \kappa^3  \langle P \rangle + N_t^{-1} 12 \times 2^{N_t} \kappa^{N_t-1}  \langle {\rm Re}\Omega \rangle\right)
+ O(\kappa^6)
\right] 
\nonumber \\ && 
- \; [T=0] 
\label{eq:e3p}
\end{eqnarray}
where $[T=0]$ is the zero temperature contribution to be subtracted for renormalization.
%The beta functions $a(d \beta /da)$ and $a(d \kappa /da)$ are the derivatives along a line of constant physics in the $(\beta, \kappa)$ plane. 
At $\kappa=0$, $(1/6)a(d \beta /da) \approx -(0.064$--$0.078)$ at $\beta=\beta_c(N_t=4)$, depending on the method to define the scale \cite{Edwards:1997xf,boyd,taro,Ejiri:1998}.  
At $\kappa=0$, $\langle {\rm Re} \Omega \rangle = 0$ due to the center $Z(3)$ symmetry, and $a(d \kappa /da) = 0$ because $\kappa=0$ is a line of constant physics. 
Neglecting terms proportional to $\kappa^3 a(d \kappa /da)$ at small $\kappa$, the latent heat is roughly evaluated as
\begin{eqnarray}
\frac{\Delta \varepsilon}{T^4} = 
\frac{\Delta (\varepsilon - 3p)}{T^4} \approx 
- 6 N_t^4\, a \frac{d \beta}{d a} \, \langle \Delta P \rangle 
\label{eq:de3p}
\end{eqnarray}
where $\Delta P$ is shown in Fig.~\ref{fig:height_peak} for $N_{\rm f}=2$.
It is found from the behavior of $\Delta P$ that the latent heat decreases as $\kappa$ increases.

\section{The cases of $2+1$-flavor and degenerate $N_{\rm f}$-flavor QCD}
\label{sec:2+1flavor}

\begin{figure}[t]
  \begin{minipage}{7cm}
   \centering
    \includegraphics[width=7cm, clip]{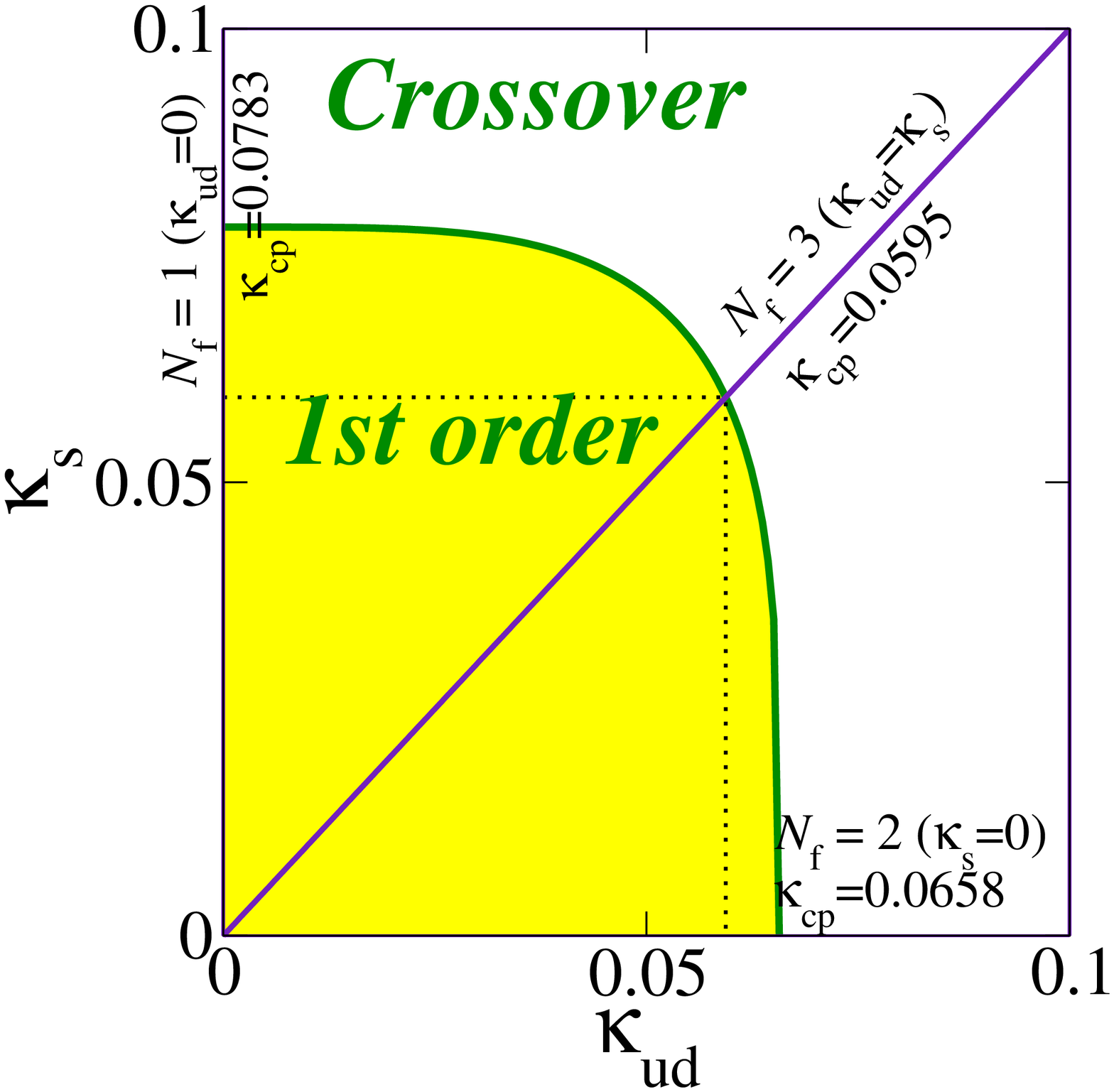}
    \caption{The phase boundary separating the first order transition region and crossover region in the $(\kappa_{\rm ud}, \kappa_{\rm s})$ plane.}
    \label{fig:Nf_Kep}
  \end{minipage} \ \ \ 
 \def\@captype{table}
 \begin{minipage}{7cm}
 \centering 
   \begin{tabular}{cc} \hline
\ \  $N_{\rm f}$  \ \     & $\kappa_{\rm cp}$  \\ \hline
        1                         &       0.0783(4)($^{+5}_{-13}$)  \\
        2                         &       0.0658(3)($^{+4}_{-11}$)  \\
        3                         &       0.0595(3)($^{+4}_{-10}$)  \\ \hline
   \end{tabular}   
   \tblcaption{$\kappa_{\rm cp}$ for $N_{\rm f}= 1$, 2 and 3 determined by the reweighting method with the lowest order hopping parameter expansion at $N_t=4$.
%$T_{\rm trans}/m_{\rm PS} \approx 0.023$ at $\kappa_{\rm cp}$ for $N_{\rm f}=2$. 
} 
   \label{tab:Kep_foreach_Nf}
  \end{minipage}
\end{figure}

The analysis can be easily generalized to an arbitrary value of $N_{\rm f}$ and also to the case of $N_{\rm f}=2+1$ QCD.
At the leading order of the hopping parameter expansion,
the contribution of quark determinants in the partition function is given by 
\begin{eqnarray}
\ln \left[ \frac{(\det M (\kappa_{\rm ud}))^2 \det M (\kappa_{\rm s})}{(\det M (0))^3} \right]  
=  288 N_{\rm site}  (2 \kappa_{\rm ud}^4 +\kappa_{\rm s}^4) P + 12 \times 2^{N_t} N_s^3 (2 \kappa_{\rm ud}^{N_t} + \kappa_{\rm s}^{N_t}) {\rm Re}\Omega 
+ \cdots
\label{eq:detm2+1}
\end{eqnarray}
for the case of $N_{\rm f}=2+1$, 
where $\kappa_{\rm ud}$ and $\kappa_{\rm s}$ are hopping parameters for light and strange quarks. 
The modification factor for the reweighting in $\kappa$  thus becomes 
\begin{eqnarray}
R(P', \kappa_{\rm ud}, \kappa_{\rm s}, 0) 
&=& e^{288 N_{\rm site} (2 \kappa_{\rm ud}^4 + \kappa_{\rm s}^4) P' } 
\nonumber \\ 
&& \times \frac{ \left\langle \delta(P'-P) \exp [
12 \times 2^{N_t} N_s^3 (2 \kappa_{\rm ud}^{N_t} + \kappa_{\rm s}^{N_t}) {\rm Re}\Omega 
+ \cdots  ] \right\rangle_{(\beta, \kappa=0)} }{
\left\langle \delta(P'-P) \right\rangle_{(\beta, \kappa=0)}}. 
\label{eq:rew2+1}
\end{eqnarray}
Since the contribution from the plaquette in this equation does not affect the second derivative of $V_{\rm eff}$, 
the difference from the case of $N_{\rm f}=2$ is just the replacement of $2\kappa^{N_t}$ by $2 \kappa_{\rm ud}^{N_t} + \kappa_{\rm s}^{N_t}$. 
Thus, the line which separates the first order phase transition and the crossover is given by 
\begin{equation}
2 (\kappa_{\rm ud})^{N_t} + (\kappa_{\rm s})^{N_t} = 2 (\kappa^{N_{\rm f}=2}_{\rm cp})^{N_t}
\end{equation}
where $N_t=4$ and $\kappa^{N_{\rm f}=2}_{\rm cp}=0.0658(3)(^{+4}_{-11})$ as determined in Sec.~\ref{sec:criticalpoint}.
We draw the line in Fig.~\ref{fig:Nf_Kep}, in which the colored region corresponds to the first order transition.
The case of degenerate $N_{\rm f}$-flavor QCD can be similarly investigated.
Results for $\kappa_{\rm cp}$ are summarized in Table \ref{tab:Kep_foreach_Nf} for $N_{\rm f} =1$, 2 and 3.
Our $\kappa_{\rm cp}$ for $N_{\rm f}=1$ is consistent with the result obtained by an effective model in Ref.~\cite{Alexandrou:1998wv}.
In the present approximation of the lowest order hopping parameter expansion at $N_t=4$, $\beta_{\rm cp}$ is independent of $N_{\rm f}$.

\section{Conclusion}
\label{sec:conclusion}

We have investigated the order of the deconfinement phase transition in the heavy quark region of QCD by calculating an effective potential $V_{\rm eff}(P)$ defined as the logarithms of the probability distribution function for the average plaquette $P$.
To study the fate of the first order deconfinement transition, $V_{\rm eff}$ has to be reliably evaluated in a wide range of $P$ covering the both phases.
Applying a reweighting method, we combine the results at five different $\beta$ points to calculate $V_{\rm eff}$ in a wide range of $P$.
We evaluate $V_{\rm eff}$ at $\kappa=0$ by simulations in SU(3) pure gauge theory, and reweight it to nonzero but small values of $\kappa$ using the leading order hopping parameter expansion on an $N_t=4$ lattice.

At $\kappa=0$, $V_{\rm eff}(P)$ show clear double-well structure in accordance with the first order deconfinement transition of the SU(3) pure gauge theory.
When we increase $\kappa$ from zero, the double-well shape becomes weaker, and eventually disappear at finite $\kappa$, indicating that the first order transition turns into a crossover at that point as suggested from the effective Z(3) model with an external magnetic field. 
We estimated the critical point $\kappa_{\rm cp}$ by examining various properties of $V_{\rm eff}$.
The results for the critical point in degenerate $N_{\rm f}$-flavor QCD as well as that in the $N_{\rm f}=2+1$ QCD are summarized in Table~\ref{tab:Kep_foreach_Nf} and Fig.~\ref{fig:Nf_Kep}.

The calculations are done in the leading order approximation of the hopping parameter expansion on an $N_t=4$ lattice.
Although the values of $\kappa_{\rm cp}$ we obtained are quite small, it is important to confirm the reliability of the leading order approximation quantitatively. 
In the next leading order, we have $O(\kappa^6)$ contributions from Wilson loops with length 6 and 
$O(\kappa^{N_t + 2})$ contributions from Polyakov loops which have $N_t + 2$ link variables including two spatial links.
To estimate the effects of them, measurement of the probability distribution function including expectation values for these operators is required.
Another important point to be studied is the continuum extrapolation by increasing $N_t$.
We leave these studies as future works.

\section*{Acknowledgments}
We would like to thank Y.~Taniguchi for discussions.
This work is in part supported by Grants-in-Aid of the Japanese Ministry of Education, Culture, Sports, Science and Technology (Nos.\  21340049, 22740168, 22840020, 20340047, 23540295) 
and by the Grant-in-Aid for Scientific Research on Innovative Areas
(Nos.\ 20105001, 20105003, 23105706).
HO is supported by the Japan Society for the Promotion
of Science for Young Scientists.

\end{document}